\documentstyle[12pt,a4]{article}
\newcommand{\ns}{\normalsize}
\newcommand{\ep}{$\rm e^+e^-$}
\newcommand{\wpm}{$\rm W^\pm$}
\newcommand{\ww}{$\rm e^+e^-\to W^+W^-$}
\newcommand{\wwz}{$\rm e^+e^-\to W^+W^-Z$}
\newcommand{\wwa}{$\rm e^+e^-\to W^+W^-\gamma$}
\newcommand{\wwzlll}{$\rm e^+e^-\to W^+_LW^-_LZ^{ }_L$}
\newcommand{\wwll}{$\rm e^+e^-\to W^+_LW^-_L$}
\newcommand{\gev}{\,\rm GeV}
\newcommand{\fb}{\,\rm fb}
\newcommand{\be}{\begin{equation}}
\newcommand{\ee}{\end{equation}}
\newcommand{\ws}{\protect\sqrt{s}}

\begin{document}
\begin{titlepage}
\thispagestyle{empty}
\newlength{\oct}
\settowidth{\oct}{December 1992\hspace{3mm}}
\begin{flushright}\begin{minipage}[t]{\oct}\raggedright BI-TP 92/59\\
LMU-92/15\\hep-ph/9212291\\December 1992\end{minipage}\\\end{flushright}
\vspace{2cm}
\renewcommand{\thefootnote}{\fnsymbol{footnote}}
\begin{center}\large{\bf
BOUNDS ON BESS MODEL  PARAMETERS\\ FROM VECTOR-BOSON PRODUCTION\\
IN $\bf e^+e^-$ COLLISIONS\footnote{Supported in part
by Deutsche Forschungsgemeinschaft, Project No.: Ko 1062/1-2}} \\ \vspace{1cm}
R. B\"onisch${ }^{1,2}$\footnote{E-Mail:
rb@hep.physik.uni-muenchen.de},
C. Grosse-Knetter${ }^{2}$\footnote{E-Mail: knetter@physf.uni-bielefeld.de} and
R. K\"ogerler${ }^{2}$ \\\vspace{1cm} \ns
${ }^{1}$Sektion Physik, Universit\"at M\"unchen, 8000 M\"unchen 2,
Germany\\\ns
${ }^{2}$Fakult\"at f\"ur Physik, Universit\"at Bielefeld,
4800 Bielefeld 1, Germany \end{center}
\renewcommand{\thefootnote}{\fnsymbol{arabic}}
\setcounter{footnote}{0}
\vspace{3.5cm}
\begin{abstract}
The BESS model is the Higgs-less alternative to the standard model of
electroweak interaction, based on nonlinearly realized spontaneous
symmetry breaking. Since it is nonrenormalizable, new couplings
(not existing in the SM) are induced at each
loop order. On the basis of the one loop induced
vector-boson self-couplings we study
the two- and three-vector-boson-production processes in $\rm e^+e^-$
collisions at $\sqrt{s}=500\rm GeV$, the expected energy of the next
$\rm e^+e^-$ linear collider (NLC). Assuming that NLC results will agree
with the SM predictions within given accuracy we identify the bounds for the
free parameters of the BESS model.
\end{abstract}
\end{titlepage}

\section{Introduction}
One of the most important topics of future particle physics
will be the investigation of the self-interactions of the electroweak
gauge bosons \wpm , Z and $\gamma$.
Since the fermionic part
of the SM is well verified by experiments performed up to
now, most alternative models
are characterized by deviations of the gauge-boson self-couplings from
the Yang--Mills type.

Presently we have only indirect empirical knowledge of these
self-interactions due to loop effects, and \ep -colliders with energies beyond
the $\rm W^+W^-$ threshold will be necessary for direct tests as well.
The most powerful tests will come from
measurements of the cross sections for
two- and three-vector-boson-production processes
in electron-positron annihilation. The first machine that is able to produce
the process
\ww\ will be LEP II, where
the available energy of $\ws\approx 190\gev$
is only slightly above the $\rm W^+W^-$ threshold.
However, LEP II will not
be very sensitive to deviations of the self-couplings from the SM values,
since these cause a violation of the gauge cancellations and lead
to differences from the SM increasing with energy. The proposed \ep -collider
working at an energy of $\ws\approx 500\gev$ and having a high
luminosity of $20\fb$ (NLC)
\cite{nlc}, however, will establish good possibilities
to discriminate
alternative models, like the BESS model, from the SM. Futhermore
it will allow the measurement of three-gauge-boson-production
processes, which are able to test the quartic self-couplings.

One principal way of testing the vector-boson self-interactions is to start
from
a Lagrangian containing the most general form of these
self-couplings which can be constructed in agreement with Lorentz
invariance. But due to the complicated structure of the general
Lagrangian and the large number of free parameters
this is a complicated procedure involving elaborate
multi-parameter fits \cite{bimoth}, especially if one investigates
not only the cubic, but also the quartic self-couplings.
In comparison it is  more effective
to test directly the sector of self-interactions
in terms of physically meaningful deviations given by (reasonable) models.

In the present paper we investigate the implications of
the structure of the vector-boson self-couplings in the
BESS (Breaking Electroweak Symmetry Strongly)
model \cite{ccdg,cvko1}. This model is based upon
a very
different mechanism for gauge-boson-mass generation, since
there are only
as many scalar fields in this theory
as needed to supply the additional degrees of freedom
for massive gauge bosons. So all scalars are unphysical would-be-Goldstone
bosons
and the model contains no physical Higgs boson. However, avoiding the Higgs
bosons necessarily implies nonlinear realization of the gauge symmetry in
the scalar sector so that the model is nonrenormalizable.
Consequently, it has to be considered as an effective
theory, which describes physics only over a restricted energy range
beyond which ``new physics'' arise. In difference to renormalizable
models, at each loop-order new couplings are induced, which do not
have the tree-level structure and cannot be removed by ordinary
counterterms (that have
the structure of the original Lagrangian).
These quantum induced interactions
have necessarily to be taken into account if the model is
taken seriously \cite{apbe}. For gauge-boson self-interactions they have
been completely calculated to one-loop level \cite{cvko2}.
Using the resulting vertices, the cross sections for two- and three-gauge-boson
production have been calculated for reasonable choices of the free parameters
of the BESS model \cite{cvgkko} with the result that the BESS model
leads to measurable differences from the standard model at NLC energies (for
these parameter choices).

Our purpose is now to determine the empirical limits which can be assigned
to the (yet free) BESS model parameters if the standard model cross section
for vector-boson production will be confirmed in NLC experiments, i. e. to
find out those values of the free parameters for which the predicted
differences of the BESS model to the SM do not exceed the
empirical error. In other words, if the BESS model is realized in
nature and the parameters lie outside the abovmentioned regions, the
NLC experiments would exhibit measurable deviations from the SM predictions.
The present analysis will also show which
physical quantities are most
sensitive to varitions of the free parameters.
Such parameter
fits have already been performed for the BESS model on the basis of processes
which get contributions from the induced femionic coupligs of the gauge bosons
\cite{boko} while here we study the implications of
the vector-boson
self-couplings, thereby taking the quantum induced contributions fully into
account.
Our investigations are based on the total cross sections for
the reactions \ww , \wwz\ and \wwa\ (production of polarized and of
unpolarized gauge bosons) and on the forward-backward asymmetry
for the two-gauge-boson-production process.

In Section 2 we give a brief overview of the BESS model and scetch the features
which are most important for our purposes. In Section 3 we explain the methods
and the phenomenological input used to perform the parameter fits. In Section 4
we present and discuss the results of these parameter fits. Section 5 contains
our conclusions.

\section{The BESS Model}
We will not present the full model here (for this see \cite{ccdg,cvko1}),
instead we give a short overview and outline the aspects which are fundamental
for the following analysis.

The BESS model is a spontaneously
broken gauge theory of electroweak interaction.
In difference to the SM, the scalars are realized by nonpolynomial
expressions and transform nonlinarly under gauge transformations. By this way
of constructing spontaneous symmetry breaking
one can avoid the introduction of physical Higgs-bosons:
there are as many scalars as massive gauge bosons. The simplest model of
this type for electroweak
interaction is the $\rm SU(2)_L\times U(1)_Y$-gauged nonlinear
$\sigma$-model \cite{apbe}, which is just the limit of the SM for infinite
Higgs mass. This model contains additional
local (``hidden'') symmetries \cite{bkuyy}, which can be made apparent by
increasing the number of unphysical scalar degrees of freedom and introducing
the gauge bosons
connected to the hidden symmetry groups \cite{ccdg,cvko1}.
The BESS model is the simplest
of these extensions of the nonlinear $\sigma$-model
with one additional $\rm SU(2)_V$
gauge symmetry and the corresponding gauge boson triplet.

The fundamental parameters of the BESS model are the following: $g,g',g''$
(the coupling constants of the gauge groups $\rm SU(2)_L,\, U(1)_Y$ and $\rm
SU(2)_V$), the vacuum expectation value $f^2$ and the relative strength of
the hidden symmetry $\lambda^2$. The (unphysical) gauge fields are
$\rm \vec{\tilde{W}},\, \tilde{Y}$ and $\rm\vec{\tilde{V}}$. They belong
to the three gauge groups, that mix to the mass eigenstates $\rm W^\pm,\,
Z,\, \gamma,\, V^\pm$ and $\rm V^0$, which are all except for the photon
massive. Low energy phenomenology demands a large coupling constant $g''$
which yields heavy V bosons.

To perform loop calculations within the nonrenormalizable BESS model one has to
introduce a cut-off $\Lambda$
for making all divergent integrals finite.
$\Lambda$ roughly represents the scale of new physics.
In calculating
loops one has to distinguish between
\begin{itemize}
\item Those cut-off dependent terms whose
structure
exists already in the
starting Lagrangian and which can therefore be treated by the usual
renormalization procedure
(addition of counterterms) and
\item The new ``induced'' couplings that reflect the nonrenormalizability
of the BESS model.
\end{itemize}
For the results of these calculations see
\cite{cvko1,cvko2,cvgkko}. The main points are:
\begin{itemize}
\item All induced couplings have a logarithmic cut-off dependence.
\item The induced fermionic couplings are suppressed by a factor of
$(m_f/M_W)^2$ and therefore negligible for all fermions except for b- and
t-quarks.
\item The induced gauge-boson self-couplings are proportional to polynomials
in the parameter $\lambda^2$, of the third power for cubic and of the fourth
power for quartic self-interactions.
\item The induced cubic self-couplings show Yang--Mills structure while the
induced quartic couplings do not.
\end{itemize}
These results show, that the most genuine feature of the BESS model (and in
fact of all theories with nonlinear symmetry realiztion), namely the existence
of quantum induced interactions, shows up most clearly in processes which get
contributions from gauge boson self-couplings.
The most prominent examples are
the various IVB-production processes in \ep -collisions. Furthermore, since
the size of the induced coupling strength grows with $\lambda^2$,
which on the other hand also governs the size of $M_V$ (the mass of the heavy
vector bosons), one expects that the influence of the induced interactions
increases with increasing $M_V$.

\section{Fit of Free Parameters}
The BESS model contains the five abovementioned free parameters, two more than
the nonlinear $\sigma$-model. These five parameters are related to
the physical coupling constants and masses as described
in \cite{ccdg,cvko1,cvgkko}. In particular, these parameters can be
reconstructed from the values of $M_W,\,
M_Z$ (gauge boson masses), $\alpha$ (electromagnetic fine structure constant),
$g/g''$ and
$\lambda^2$. While the first three of these are given by experiments at present
energies, the latter two are yet unspecified.
Future high energy experiments at NLC will allow a specifcation of these two
free parameters within experimental errors.

Our question is now which numerical values of the free parameters $g/g''$ and
$\lambda^2$ will yield measurable differences to the SM predictions at NLC for
various physical quantities (cross sections and asymmetries for two- and
three-gauge-boson-production processes) and which will not.
So on the one hand we derive bounds on the two free parameters for the
case that NLC confirms the SM predictions and on the other hand we determine
the
regions in the ($\lambda^2,g/g''$) plane where the BESS model will be be
empirically distinguishable from the SM and which
physical quantities to be measured at NLC are especially sensitive to the BESS
model.
In addition, we will demonstrate which quantities
are more sensitive to $g/g''$ and which are more sensitive to $
\lambda^2$.

As an input for our calculations we need, as mentioned above, the gauge boson
masses $M_W$ and $M_Z$ and the electromagnetic fine structure constant
$\alpha$. The latter is taken as
\be \alpha=\frac{1}{127}\, ,\ee
which is the value of the runnig coupling constant in the NLC region.
A further needed parameter
is the cut-off $\Lambda$ that is used to calculate the induced couplings.
We choose
\be \Lambda=5\rm\, TeV\footnote{Because of the logarithmic cut-off dependence a
slightly different choice of $\Lambda$ does not significantly
influence the results.}\, .\ee
Based upon these inputs and using freely
chosen values for $g/g''$ and $\lambda^2$, it
is possible to determine the cross sections, asymmetries, etc. and to compare
them
with the SM results\footnote{For the SM reference values we neglect all Higgs
effects (which contribute to \wwz ),
since for a Higgs mass $M_H > \ws-M_Z$
these contributions are negligible.
Otherwise the Higgs boson would be found at
NLC in the reaction $\rm e^+e^-\to ZH$ and the BESS model would be ruled out
anyway.}. The cross sections for two-gauge-boson production are
calculated completely analytically using the usual trace techniques, while for
the calculation of three-gauge-boson-production processes we use the
same numerical techniques as in ref.~\cite{bahaph}.
In agreement with \cite{bahaph} we impose the following cuts
on transverse-momentum $P_{T,\gamma}$ and pseudorapidity $\eta_\gamma$ of
the photon produced in \wwa :
\be P_{T,\gamma}>20\gev\, ,\qquad |\eta_\gamma|<2\, .\ee
All calculations are performed at the NLC energy of
\be \ws=500\gev\, .\ee

To perform the parameter fit one has to know the expected experimental
accuracy (sum of the statistical and the systematical error).
The
statistical error is predicted from the event rate,
assuming an integrated luminosity of $20\fb^{-1}$ per year
and taking into account a reduction factor (branching fraction and acceptence
correction) of 0.3 for \ww\ \cite{frmaseze} and \wwa\ and of 0.2 for \wwz\
\cite{bahaph}. We determine
the statistical error to 90\% confidence level.
We consider both
a ``conservative'' accuracy, meaning one year data collecting time
plus a systematical error of 1.5 \%
and an ``ideal'' one, assuming three years of running plus
 a systematical error of
1.2 \% for \ww . For three gauge-boson production we assume a systematical
error of 2\% and three years of collecting data.

For the process \ww\ (unpolarized gauge bosons)
the  total cross section is $8\,\rm pb$, and so there
will be a large number of events and the
systematical error dominates, but if one studies cross sections for the
production of polarized gauge bosons, partial cross sections or
three-gauge-boson-production processes, the event rate is considerably
smaller (for \wwz\
the total cross section is $47\fb$) and the statistical error becomes more
important. So larger deviations to the SM in these cases do not
neccesarily imply stricter bounds on the free parameters. That is why we
did not consider bounds obtained after only one year of run for
three-gauge-boson production;
the error would be too large to get useful results.

\section{Discussion of the Results}
Out of the observables that were calculated in ref.~\cite{cvgkko} we find
the following ones to be most interesting for deriving bounds to the BESS
model:
\begin{itemize}
\item The total cross section for production of two unpolarized Ws, i.~e.
\ww .
\item The total cross section for production of two longitudinally polarized
Ws, i.e. $\rm e^+e^- \rightarrow W^+_L W^-_L$.
\item The forward-backward asymmetry of $\rm W^+W^-$  production.
\item The total cross sections for the three-gauge-boson-production processes
\wwz\ and \wwa\ (production of unpolarized gauge bosons).
\item The total cross section for production of three longitudinally polarized
gauge bosons \wwzlll .
\end{itemize}

Fig.~\ref{ww} shows the bounds (in the ($\lambda^2$, $g/g^{\prime\prime}$)
plane)
stemming from production of an unpolarized W pair. In all figures
the  shaded areas  below the curves are the allowed regions,
the  dashed line corrsponds to the conservative, the solid
one to ideal precision,
as characterized in the last section.
{}From these data $\lambda^2$ is restricted to be lower than 10 and $g/g''$ to
be
lower than 0.12 for the conservative
measurement. The bounds obtained from the ideal measurement are only
slightly stricter because the statistical error is small anyhow.
Production of two longitudinally polarized gauge bosons
yields stronger limits (Fig.~\ref{wwll}),
since the greatest deviations of the BESS results from the SM predictions
occur in
the production of purely longitudinal gauge bosons due to the violation of the
gauge cancellations caused by the induced couplings. Here, accuracy
increases much for the ideal measurement, i.~e.\ for a longer time of
collecting data since this reduces the statistical error arising as a
consequence of
the small cross-section for this channel.
Fig.~\ref{afb} shows the bounds stemming from the forward-backward asymmetry.
The conservative measurement does not yield any interesting bounds, especially
with respect to $g/g''$. The ideal mesurement gives bounds on $\lambda^2$
comparable to those from $\sigma_{tot}$ but worse limits on $g/g''$.

Figs.~\ref{wwz} and \ref{wwa} exhibit the bounds that result from the total
cross sections
for \wwz\ and \wwa\ (unpolarized gauge bosons), respectively. These processes
are of big importance, since they allow a direct test of four-gauge-boson
self-interactions (together with three-boson ones). Because of the stronger
dependence of the induced quartic self-couplings on $\lambda^2$ there are
larger deviations from the SM (at least for higher $\lambda^2$-values) within
these processes \cite{cvgkko}. Unfortunately however, the corresponding
cross sections are much smaller than the two gauge-boson-production cross
sections. Therefore,
the statistical accuracy to be obtained at NLC is worse and
we get bounds which are not much better than those stemming from
two-gauge-boson
production\footnote{The results should be compared to those of
the ideal measurement in Fig.~1, which are based on three years' data, too.}.
Fig.~\ref{wwz} shows that $g/g''$ can be
restricted to be lower than 0.1 and $\lambda^2$ to be lower then 10, if the SM
is confirmed at NLC. \wwa\ yields slightly stricter bounds on
both parameters (Fig.~\ref{wwa}) due to the larger cross section of $153\fb$,
i.~e., the smaller statistical error.

We also have analyzed
the bounds arising from the reaction \wwzlll\ (Fig.~\ref{wwzlll}) with
longitudinal final gauge bosons.
Unfortunately
the cross section is so small ($0.5\fb$) and the statistical error becomes
so large that the effect of larger deviations is again compensated.
This
reaction therefore does not yield stricter bounds than production of
unpolarized gauge
bosons at the NLC (in difference to W-pair production).
However, comparing Figs.~\ref{wwz} and \ref{wwzlll}, one can see
that production of unpolarized gauge bosons is more sensitve to $g/g''$, while
production of longitudinally polarized gauge bosons is more sensitive to
$\lambda^2$, i. e. to the strength of the induced couplings.

Because of the insufficient
statistics caused by the small cross sections an analysis
of partial cross sections
would not yield stricter bounds. The same is true for the reaction $\rm e^+e^-
\to ZZZ$ with a tiny total cross section of only $1\fb$. So we did not apply
our analysis to these quantities.

The resulting
bounds on BESS model parameters  should be compared to those which are
obtained from fermionic processes measured at LEP~I
(\cite{boko} Figs. 8 - 12). We see that gauge
boson production will yield stricter bounds in most cases. In particular,
higher values of $\lambda^2$, which are allowed by most data from
fermionic processes, can be excluded. This is not surprising, since $\lambda^2$
governs the strength of the induced couplings and these do not contribute very
much to the fermionic sector of the theory but to the bosonic
self-interactions. Only the fermionic process $\rm e^+e^-\to b\bar{b}$
at LEP I energies yields bounds comparable to the ones obtained here or even
better.\footnote{However the small errors for $A_{FB}$ and $\sin^2\theta_W$
assumed  in \cite{boko} are not reached by LEP I data so that the bounds from
$\rm e^+e^-\to b \bar{b}$ are not so strict as given there.}

\section{Conclusion}
We summarize our main results:
\begin{itemize}
\item The strictest bounds
to the free BESS-Model parameters arise from the total
cross sections while the forward-backward assymetry
for \ww\ is less sensitive to $g/g''$.
\item The processes \ww\ and \wwa\ yield better bounds than \wwz\ due
to better statistics.
\item Production of longitudinally polarized gauge bosons yield better
limits for two-gauge-boson production. For three-gauge-boson production
the results are comparable to those from
production of unpolarized gauge bosons but more sensitive to
$\lambda^2$.
\item Measurements of gauge-boson-production at NLC will
improve the information on the
evidence of the model with respect to fermionic processes measured at LEP~I
\cite{boko}.
This is due to the induced self-couplings of the vector bosons, whereas the
induced fermionic interactions are very small\footnote{See, in
comparison, the treatment of \cite{ruj}, which represents a principally
different theoretical acces to nonstandard vector-boson self-couplings.
There gauge invariant extra terms with {\em linearly\/} realized symmetry are
added to the SM Lagrangian which yield anomalous couplings on {\em
tree-level\/}.
In the BESS model gauge symmetry is realized {\em nonlinearly\/} and anomalous
couplings are {\em loop implied\/}.
However, both acceses yield  the result that anomalous self-interactions
(within the two different formalisms)
are not restricted very well by LEP~I measurements.}.

\end{itemize}
The above discussion has shown that, if the SM will be
confirmed at NLC, the free BESS model parameters will be restricted to a quite
narrow region. For a wide parameter range the structures
of the BESS model (heavy vector bosons, mixing between light and heavy bosons,
new induced couplings due to nonrenormalizability) will yield measurable
effects on two- and three-gauge-boson-production processes
at NLC. So there is hope that indications of the BESS model
(if this is realized in nature) may be found at NLC.
However one has to collect data over a longer period of
time to get useful results due to limited statistics.

\section*{Acknowledgement}
We thank G. Cveti\v{c} for discussing the analytic parametrization of the
two-boson-production quantities.

\section*{Figure Captions}
\newcounter{fig}
\newlength{\fig}
\newlength{\numnum}
\settowidth{\numnum}{\bf 1}
\settowidth{\fig}{\bf Figure 1:}
\begin{list}{\bf Figure \makebox[\numnum][r]{\arabic{fig}}:}{\usecounter{fig}
\labelwidth\fig \leftmargin\labelwidth \addtolength\leftmargin\labelsep}
\item \label{ww} Bounds from the total cross section for \ww\ (unpolarized)
\item \label{wwll} Bounds from the total cross section for \wwll\
(longitudinally polarized)
\item \label{afb} Bounds from the forward-backward assymmetry for \ww\
(unpolarized)
\item \label{wwz}Bounds from the total cross section for \wwz\ (unpolarized)
\item \label{wwa}Bounds from the total cross section for \wwa\ (unpolarized)
\item \label{wwzlll}Bounds from the total cross section for
\wwzlll\ (longitudinally polarized)
\end{list}
\end{document}